\pdfoutput=1
\documentclass[12pt]{article}

\usepackage{jcappub}
\usepackage{amsmath}
\usepackage{amssymb}
\usepackage{braket}
\usepackage{graphicx}
\usepackage{mathrsfs}
\usepackage{subcaption}
\usepackage{empheq}
\usepackage{appendix}
\usepackage{natbib}
\usepackage[dvipsnames]{xcolor}
\usepackage{float}
\usepackage{indentfirst}
\usepackage[top=1in, bottom=0.4in, left=0.8in, right=0.8in, includefoot]{geometry}

\title{Constant-roll in the Palatini--$R^2$ models}
\author[a,b]{Ignatios Antoniadis,}
\author[c]{Angelos Lykkas,}
\author[c,b]{Kyriakos Tamvakis}

\affiliation[a]{Laboratoire de Physique The\'orique et Hautes Energies - LPTHE, Sorbonne Universit\'e, CNRS, 4 Place Jussieu, 75005 Paris, France}
\affiliation[b]{Albert Einstein Center, Institute of Theoretical Physics, University of Bern, Sidlerstrasse 5, CH-3012, Bern, Switzerland}
\affiliation[c]{Physics Department, University of Ioannina, GR--45110 Ioannina, Greece}

\emailAdd{antoniad@lpthe.jussieu.fr}
\emailAdd{a.lykkas@uoi.gr}
\emailAdd{tamvakis@uoi.gr}

\abstract{We consider models of a scalar field coupled to quadratic $R\!+\!R^2$ gravity in the framework of the Palatini formulation. The resulting Einstein-frame generalized $k$-inflation effective theory is analyzed assuming that the constant-roll condition holds. We focus on a quartic self-interaction potential, a case of particular appeal modelling Higgs inflation, considering the cases of minimal and non-minimal coupling of the inflaton to gravity. For an appropriate range of the model parameters in the large field domain the obtained values for the inflationary observables are found in agreement with current observations.}

\notoc
\newcommand{\be}{\begin{equation}}
\newcommand{\ee}{\end{equation}}

\begin{document}

\maketitle

\clearpage

\vspace*{3.5cm}
\section{Introduction}

The theory of cosmic inflation~\cite{Starobinsky1980, Guth1981, Sato1981, Linde1982, Albrecht1982, Linde1983a, Lyth1999}, i.e. a period of quasi--de Sitter expansion of the Universe during the first instants after its birth, was initially motivated by its solution to the problems of traditional Big Bang cosmology, like the observed flatness and large scale CMB temperature uniformity. Nevertheless, its present appeal is that it provides a mechanism through which the tiny primordial inhomogeneities, arising as quantum fluctuations, are allowed to grow and become classical at superhorizon scales~\cite{Starobinsky1979, Mukhanov1981b, Hawking1982a, Hawking1983, Starobinsky1982a, Guth1982a}. Present experimental efforts focused on the CMB~\cite{Akrami2018, Ade2018}, strongly constrain the inflationary spectrum with an increasing accuracy, having narrowed considerably the range of viable inflationary models. Although the power spectrum of scalar perturbations is very nearly scale-invariant and Gaussian, and this agrees with inflationary models in their simplest implementation, in the medium-term future non-Gaussianities might be uncovered through the increasing accuracy of observations. The scalar degree of freedom (inflaton) employed in these models can either be a fundamental scalar field or can arise as an effective scalar degree of freedom incorporated in gravity itself. The latter possibility is realized in the so-called modified gravity models (see for a review~\cite{DeFelice2010, Sotiriou2010, Capozziello2011, Clifton2012, Odintsov2016}) and in particular in the Starobinsky model\cite{Starobinsky1980}, which is persistently in agreement with observations, regarding its inflationary spectrum. In fact, the Starobinsky model, as well as any theory of gravity with an action $\int\,d^4x\,\sqrt{-g}\,f(R)$, can be reformulated as a scalar--tensor theory of gravity with a non-minimal coupling of the effective scalar degree of freedom to the Ricci scalar.

Although known for sometime, the so-called first order or Palatini formulation of gravity~\cite{Palatini:1919} has recently received considerable attention~\cite{Bauer2008, Borunda2008, Bauer2011, Tamanini2011, Enqvist2012, Borowiec2012, Racioppi2017, Rasanen2017, Fu2017, Stachowski2017, Azri2017, Enckell2018, Enckell2018a, Jaerv2018, Rasanen2018, Antoniadis2018, Rasanen2018a, Antoniadis2019b, Racioppi2019, Shimada2019, Jinno2019, Tenkanen2019, Rubio2019, Gialamas2019, Tenkanen2020, Shaposhnikov2020, Tenkanen2020a, Shaposhnikov2020a, Lloyd-Stubbs2020}, since, in the case of scalar fields non-minimally coupled to gravity, it leads to different predictions than the standard (dubbed as metric) formulation. In the Palatini formulation the metric $g_{ \mu\nu}$ and the connection $\Gamma_{ \mu\nu}^{\rho}$ are treated as independent variables. Although within GR the Palatini formulation is entirely equivalent to the standard metric formulation, its application in theories containing $f(\phi)R$-type non-minimal couplings, leads to different results. For example, the Starobinsky model within the Palatini formulation does not lead to any propagating scalar degree of freedom, in contrast to its metric formulation. Quadratic gravity with an $R^2$ term combined with a fundamental scalar non-minimally coupled through $\xi\,\phi^2\,R$, considered in the Palatini framework, leads to a generalization of a $k$-inflation-type theory~\cite{Armendariz-Picon1999, Chiba2000, Armendariz-Picon2000, Armendariz-Picon2001, Chiba2002, Malquarti2003, Malquarti2003a, Chimento2004, Chimento2004a, Scherrer2004, Aguirregabiria2004, Armendariz-Picon2005, Abramo2006, Rendall2006, Bruneton2007, Putter2007, Babichev2008, Matsumoto2010, Deffayet2011, Unnikrishnan2012, Li2012}, featuring additional kinetic terms $A(\phi)(\nabla\phi)^2+B(\phi)(\nabla\phi)^4$ with field-dependent coefficients. In this article we investigate the phenomenological features of such models departing from the usual slow-roll approximation, where the second derivative in the inflaton equation of motion is neglected, and assuming that the constant-roll condition~\cite{Motohashi2015a, Motohashi2017a, Odintsov2017, Odintsov2017a, Nojiri2017, Motohashi2017b, Gao2017a, Oikonomou2017, Odintsov2017c, Oikonomou2017a, Cicciarella2018a, Anguelova2017, Karam2018c, Yi2017, Gao2019, Odintsov2020} holds true, meaning that $\ddot{\phi}/H\dot{\phi}\approx$ constant. One of the features of constant-roll is its association with non-Gaussianities in the CMB spectrum, in contrast for example to the slow-roll treatment of $k$-inflation models where non-Gaussianities are small.

In the present article we reconsider the model of a scalar field coupled to gravity in the presence of an $R^2$ term in the framework of the Palatini formalism. We investigate the inflationary dynamics assuming that the constant-roll condition is valid and focus on the case of a quartic self-interaction potential for the scalar field, a potential modelling the particularly appealing case of Higgs inflation~\cite{Bezrukov2008a, DeSimone2009, Bezrukov2009a, Barbon2009, Barvinsky2009, Lerner2010a, Bezrukov2011, Kamada2012, Bezrukov2013, Bezrukov2014, Allison2014, Hamada2014, George2014, Salvio2015, Hamada2015, Calmet2016, Artymowski2016, Rubio2018, Enckell2018}. We find that the Einstein frame Lagrangian of the model 
\begin{equation}\label{LAGRA}
 {\cal{L}}\,=\,A(\phi)(\nabla\phi)^2+B(\phi)(\nabla\phi)^4-U(\phi),
\end{equation}
corresponds to a generalized $k$-inflation model with a significant contribution of the higher order kinetic terms to the predictions concerning the inflationary observables. All predicted observables are in agreement with the limits set by current observations for a reasonable range of the model parameters, while the inflationary dynamics takes place in the large field domain. This is in contrast to the slow-roll case where this was achieved only at the expense of an unusually large amount of $e$-folds~\cite{Antoniadis2019b}. Both the cases of minimal as well as non-minimal coupling of the scalar to gravity lead to acceptable results.

The paper is organized as: In the following section we formulate the theory of quadratic gravity in terms of an auxiliary field $\chi$ in the presence of a fundamental scalar field $\phi$ with a possible non-minimal coupling and self-interaction potential $V(\phi)$. Under a Weyl rescaling of the metric and after solving the constraint equation of the auxiliary field, we arrive at the Einstein frame with a Lagrangian of the form of \eqref{LAGRA}. Assuming a flat FRW background metric, we obtain the equations of motion of the theory. Then, in section \ref{s3}, we set up the inflationary parameters and observables under the assumption of the constant-roll condition. In section \ref{s4}, we focus on the quartic model and study its predictions for the primordial tilt, tensor-to-scalar ratio and power spectrum for the cases of minimal and non-minimal coupling to gravity. Finally, in the last section we summarize our conclusions.

\section{Quadratic gravity in the Palatini formalism}

Consider a scalar field $\phi$ with a self-interaction potential $V(\phi)$,  that is coupled non-minimally to gravity through a term $\xi\phi^2R/2$ in the presence of an $\alpha R^2/4$ term. We are anticipating that these terms are generated by radiative corrections. The interactions are parametrized in terms of the dimensionless parameters $\xi$ and $\alpha$. The resulting action
\begin{equation}
    \mathcal{S}=\int\!\mathrm{d}^4x\,\sqrt{-g}\left\{\frac{1}{2}(M_P^2+\xi\phi^2)R+\frac{\alpha}{4}R^2-\frac{1}{2}(\nabla\phi)^2-V(\phi)\right\},
\end{equation}
 can be equivalently expressed in the scalar representation, reading
\begin{equation}
    \mathcal{S}=\int\!\mathrm{d}^4x\,\sqrt{-g}\left\{\frac{1}{2}(M_P^2+\xi\phi^2+\alpha\chi^2)R-\frac{1}{2}(\nabla\phi)^2-V(\phi)-\frac{\alpha}{4}\chi^4\right\},
\end{equation}
where $\chi$ is an auxiliary scalar, satisfying $\chi^2\!=\!R$ on-shell. Going to the Einstein frame via the Weyl rescaling\footnote{Hereafter we assume Planck units, i.e. $M_P^2\equiv 1$.}
\begin{equation}
    \bar{g}_{\mu\nu}=(1+\xi\phi^2+\alpha\chi^2)\,g_{\mu\nu}
\end{equation}
and dropping the bars, we get
\begin{equation}\label{EFact}
    \mathcal{S}=\int\!\mathrm{d}^4x\,\sqrt{-g}\left\{\frac{1}{2}R-\frac{1}{2}\,\frac{(\nabla\phi)^2}{(1+\xi\phi^2+\alpha\chi^2)}-\frac{V+\frac{\alpha}{4}\chi^4}{(1+\xi\phi^2+\alpha\chi^2)^2}\right\}.
\end{equation}
Note that, since we are working within the Palatini framework, the Ricci scalar $R=g^{ \mu\nu}R_{ \mu\nu}(\Gamma)$ is only rescaled multiplicatively and no derivatives of $\phi$ and $\chi$ arise, the latter keeping its auxiliary field status. Solving the constraint equation $\delta_\chi\mathcal{S}\!=\!0$ for $\chi^2$, we obtain
\begin{equation}
    \chi^2=\frac{4V+(1+\xi\phi^2)(\nabla\phi)^2}{(1+\xi\phi^2)-\alpha(\nabla\phi)^2}
\end{equation}
and substituting back into \eqref{EFact} we arrive at the action
\begin{equation}
    \mathcal{S}=\int\!\mathrm{d}^4x\,\sqrt{-g}\left(\frac{1}{2}R+\mathcal{L}(\phi,X)\right),
\end{equation}
expressed in terms of an effective Lagrangian
\begin{equation}
    \mathcal{L}(\phi,X) \equiv A(\phi)X+B(\phi)X^2-U(\phi)\,,
\end{equation}
where
\begin{equation}
 X\,\equiv\,\frac{1}{2}(\nabla\phi)^2\,\,\,\,\,{\text{and}}\,\,\,\,\,\,\left\{\begin{array}{l}
A(\phi)\,\equiv\,-\left(1+\xi\phi^2+4\alpha\frac{V(\phi)}{(1+\xi\phi^2)}\right)^{-1}\\
\,\\
B(\phi)\,\equiv\,\alpha\left(\,(1+\xi\phi^2)^2+4\alpha V(\phi)\right)^{-1}\,=\,-\frac{\alpha A(\phi)}{(1+\xi\phi^2)}\\
\,\\
U(\phi)\,\equiv\,V(\phi)\,\left(\,(1+\xi\phi^2)^2+4\alpha V(\phi)\right)^{-1}
\end{array}\right. 
\end{equation}
Note that ${\cal{L}}$ features up to quartic kinetic terms with field-depended coefficients, belonging to a generalized class of $k$-inflation models.

The energy-momentum tensor corresponding to the source field $\phi$, governed by $\mathcal{L}$, is
\begin{equation}
    T_{\mu\nu}\equiv \frac{2}{\sqrt{-g}}\,\frac{\delta\mathcal{S}}{\delta g^{\mu\nu}}=-\frac{\partial\mathcal{L}}{\partial X}\,\left(\nabla_{\mu}\phi\right)\left(\,\nabla_{\nu}\phi\right)+g_{\mu\nu}\mathcal{L}
\end{equation}
or
\begin{equation}
T_{ \mu\nu}\,=\,-(A+2BX)\left(\nabla_{ \mu}\phi\right)\left(\nabla_{ \nu}\phi\right)+g_{ \mu\nu}(AX+BX^2-U)\,.
\end{equation}
Then, assuming that the scalar field is spatially homogeneous, depending only on time, we can obtain the energy density $\rho=T_{00}$ and the pressure as $T_{ij}=p\,g_{ij}=\mathcal{L}\,g_{ij}$, i.e.:
\begin{align}
    \rho&=A(\phi)X+3B(\phi)X^2+U(\phi),\label{g00}\\
    p&=A(\phi)X+B(\phi)X^2-U(\phi).
\end{align}
Assuming a spatially flat FRW metric, the equations of motion are:
\begin{align}
    &3H^2=\rho,\\
    &\dot{\rho}+3H(\rho+p)=0,
\end{align}
where the dot denotes derivative with respect to the cosmic time $t$. These equations combined give the following equation
\begin{equation}
    2\dot{H}+3H^2=-p.
\end{equation}
Finally, the scalar field equation of motion is
\be
    \ddot{\phi}(A+6BX)+3H\dot{\phi}(A+2BX)-A'X-3B'X^2=U'.{\label{EQOFMO}}
\ee
Hereafter, we denote with a prime the derivative with respect to $\phi$.

\section{Constant-Roll Inflation}\label{s3}

The field $\phi$, being the sole scalar degree of freedom, is our candidate for the inflaton. We shall assume that the constant-roll condition
\begin{equation}
    \ddot{\phi}=\beta H\dot{\phi},
\end{equation}
is satisfied, $\beta$ being a constant parameter. This condition approaches the slow-roll condition where $\ddot{\phi}\simeq0$ in the limit $\beta\ll 1$.  Let us also introduce the following slow-roll parameters (SRP)~\cite{Odintsov2020}
\begin{equation}\label{srps}
    \epsilon_1=-\frac{\dot{H}}{H^2},\qquad\epsilon_2=-\frac{\ddot{\phi}}{H\dot{\phi}},\qquad\epsilon_3=\frac{\dot{F}}{2HF},\qquad\epsilon_4=\frac{\dot{E}}{2HE},
\end{equation}
where
\begin{equation}
    F=\frac{\partial \mathcal{L}}{\partial R}, \qquad E=-\frac{F}{2X}\left(X\,\frac{\partial\mathcal{L}}{\partial X}+2X^2\,\frac{\partial^2\mathcal{L}}{\partial X^2}\right).
\end{equation}
We shall also assume that $\dot{\phi}^2/2\ll U(\phi)$, at least during the initial stages of inflation, so that the parameters defined above also make sense. Later on, we shall check the smallness of the SRPs. Note that the constant-roll condition fixes $\epsilon_2=-\beta$. Also, in our case, being in the Einstein frame where $F=1/2$, we have just $\epsilon_3=0$. In terms of the SRPs we may express observable quantities like the scalar spectral index (primordial tilt) $n_s$ and the tensor-to-scalar ratio $r$ as~\cite{Noh2001a, Hwang2002, Hwang2005}
\begin{align}
    n_s&=1-2\,\frac{2\epsilon_1-\epsilon_2-\epsilon_3+\epsilon_4}{1-\epsilon_1},\\
    r&=4\,|\epsilon_1|\,C_s,
\end{align}
where $C_s$ is the sound wave speed of primordial perturbations, given by
\begin{equation}
    C_s^2=\frac{\mathcal{L}_X}{\mathcal{L}_X+2X\mathcal{L}_{XX}}=\frac{A(\phi)-B(\phi)\dot{\phi}^2}{A(\phi)-3B(\phi)\dot{\phi}^2}\,=\,\frac{1+\xi\phi^2+\alpha\dot{\phi}^2}{1+\xi\phi^2+3\alpha\dot{\phi}^2},
\end{equation}
being, as it should, $0<{C_s}^2<1$. The related scalar power spectrum, to first order in the SRPs, is given by~\cite{Lyth1999}
\begin{equation}\label{powerspectrum}
P_S\,\approx\,\frac{H^2}{8\pi^2\epsilon_1}\,=\,\frac{1}{72\pi^2}\frac{\left(AX+3BX^2+U\right)^2}{\left(AX+2BX^2\right)}\,,
\end{equation}
which, evaluated at horizon crossing, should yield the observed $P_S\approx 2.1\times 10^{-9}$~\cite{Akrami2018, Ade2018}.

Aiming at obtaining $\dot{\phi}$ from the equation of motion \eqref{EQOFMO}, we insert the constant-roll condition and obtain
\begin{equation}\label{EQONO-1}
\dot{\phi}H\left[\,(\beta+3)A+6(\beta+1)BX\right]\,-A'X-3B'X^2\,=\,U'\,.
\end{equation}
Next, we obtain an approximate expression for $H$ as
\begin{equation}
H\,=\,\sqrt{\frac{U}{3}\left(1-A\left(\frac{\dot{\phi}^2}{2U}\right)\,+\,3B\left(\frac{\dot{\phi}^2}{2U}\right)^2\right)}\,\approx\,\sqrt{\frac{U}{3}}\, \left(1-\frac{A}{4U}\,\dot{\phi}^2+\frac{1}{8}\left(\frac{3B}{U}-\frac{A^2}{4U^2}\right)\dot{\phi}^4\right),
\end{equation}
where we kept terms up to $\mathcal{O}\left((\dot{\phi}^2/2U)^2\right)$. Substituting this expression in \eqref{EQONO-1}, we obtain
\begin{equation}
    \sqrt{\frac{U}{3}}\,\left(3B(\beta+1)+\frac{A^2}{4U}(\beta+3)\right)\dot{\phi}^3-\frac{1}{2}A'\dot{\phi}^2-A\sqrt{\frac{U}{3}}(\beta+3)\dot{\phi}+U'=0,\label{dotPeq}
\end{equation}
keeping only terms up to $\mathcal{O}(\dot{\phi}^3)$.\footnote{It is also possible, keeping terms up to $\dot{\phi}^4$ to solve the corresponding quartic algebraic equation  but the smallness of $\dot{\phi}$ and the negligible effect do not justify the extra complication.} This is an algebraic equation in terms of $\dot{\phi}$. It can be re-expressed as
\begin{equation}
    x^3+\nu_1\,x+\nu_0=0,\,\,\,\,\,\,\left(x\,\equiv\,\dot{\phi}-\frac{A'}{6\gamma}\,\right)\,,
\end{equation}
where
\begin{align}
    \gamma&=\sqrt{\frac{U}{3}}\,\left(3B(\beta+1)+\frac{A^2}{4U}(\beta+3)\right),\\
    \nu_1&=-\frac{1}{\gamma}\left[(\beta+3)A\sqrt{\frac{U}{3}}+\frac{(A')^2}{12\gamma}\right],\\
    \nu_0&=\frac{1}{\gamma}\left[U'-A(\beta+3)\,\frac{A'}{6\gamma}\sqrt{\frac{U}{3}}-\frac{(A')^3}{108\gamma^2}\right]
\end{align}
The real solution to the above equation reads:
\begin{equation}\label{xsol}
    x=\frac{\left(-9\nu_0+\sqrt{3}\sqrt{4{\nu_1}^2+27{\nu_0}^2}\right)^{1/3}}{2^{1/3}\,3^{2/3}}-\frac{\left(\frac{2}{3}\right)^{1/3}\nu_1}{\left(-9\nu_0+\sqrt{3}\sqrt{4{\nu_1}^2+27{\nu_0}^2}\right)^{1/3}}.
\end{equation}
Therefore, by means of the above analysis we have a solution for $\dot{\phi}[\phi]$ in terms of $\phi$ which can be substituted into all our expressions of the SRPs and spectral indices, including the integral expression for the number of $e$-folds
\begin{equation}
N\,=\,\int_{\phi_i}^{\phi_f}\frac{d\phi}{\dot{\phi}}H\,=\,\frac{1}{\sqrt{3}}\int_{\phi_i}^{\phi_f}\frac{d\phi}{\dot{\phi}}\sqrt{AX+3BX^2+U}\,.
\end{equation}

\section{Models}\label{s4}

Since $\phi$ is a fundamental scalar field, interacting with the rest of matter\footnote{These interactions should become important at the stage of reheating.}, its self-interaction potential, from a particle physics point of view, could be restricted to a renormalizable form $V(\phi)=m^2\phi^2/2+\lambda\phi^4/4$, which in the large field domain is approximated by a quartic potential $\lambda\phi^4/4$, although, admittedly, higher order terms could not be ruled out. This form of the potential is particularly appealing since it  corresponds to the inflationary dynamics of the Higgs potential $V(H)=\lambda\,(|H|^2-v^2/2)^2$ for $|H|\gg v$, driven by a nonminimal coupling with gravity of the form $\xi|H|^2$ for large values of $\xi$.\footnote{Higgs inflation has been studied in the metric formulation of gravity with the presence of an $R^2$ in \cite{Wang2017a, Ema2017a, He2018a, Enckell2018b}. Note that such a term, considered in the metric framework, would convert the dynamics into a two-field problem, since the inflaton would be a combination of the Higgs and the scalar component of gravity. In contrast, in the Palatini formulation the presence of an $R^2$ does not introduce any additional scalar.}

\vspace*{0.75cm}
{\textbf{Minimally coupled quartic potential}}\footnote{Note that, within the framework of the Palatini formalism in the presence of $R^2$ term, a non-minimal coupling of the inflaton can be Weyl-transformed away at the expense of a non-canonical kinetic term and a rescaling of the potential, since $\sqrt{-g}\,R^2$ is Weyl-invariant.} This corresponds to setting $\xi=0$. In this case we have the relations
\begin{equation}
    A=-(1+4\alpha V)^{-1},\qquad B=-\alpha A,\qquad U=-V\,A
\end{equation}
and
\begin{equation}
    A'=4\alpha A^2 V',\qquad B'=-4\alpha^2A^2V', \qquad U'=-V'A^2
\end{equation}
which simplify the general expressions for the SRPs, defined by \eqref{srps}, to
\begin{align}
    \epsilon_1&=3\,\frac{AX+2BX^2}{AX+3BX^2+U}=3\,\frac{\dot{\phi}^2+\alpha\dot{\phi}^4}{\dot{\phi}^2+\frac{3}{2}\alpha\dot{\phi}^4+2V},\\
    \epsilon_2&=-\beta,\\
    \epsilon_3&=0,\\
    \epsilon_4&=\frac{\sqrt{3}}{2}\,\frac{\dot{\phi}(A'+6B'X)+12\beta BHX}{(A+6BX)\sqrt{AX+3BX^2+U}}=\left(\frac{3\alpha\beta\dot{\phi}^2}{1+3\alpha\dot{\phi}^2}-\frac{2\sqrt{3}\,\alpha}{\sqrt{1+4\alpha V}}\,\frac{V'\,\dot{\phi}}{\sqrt{\frac{1}{2}\dot{\phi}^2+\frac{3}{4}\alpha\dot{\phi}^4+V}}\right).
\end{align}
and the scalar power spectrum as
\begin{equation}
    P_S=\frac{1}{72\pi^2}\,\frac{\left(AX+3BX^2+U\right)^2}{X(A+2BX)}\,=\,\frac{1}{36\pi^2}\frac{\left(\frac{1}{2}\dot{\phi}^2+\frac{3}{4}\alpha\dot{\phi}^4+V\right)^2}{\dot{\phi}^2\left(1+\alpha\dot{\phi}^2\right)\left(1+4\alpha V\right)}\,.
\end{equation}

In what follows we substitute the real solution of $\dot{\phi}[\phi]$ in the expressions of the slow-roll parameters. We calculate the field value $\phi_f$ at the end of inflation, by demanding $\epsilon_1(\phi_f)=1$. Finally, the start of inflation is calculated by allowing for a range of $(50-60)$ e-folds to pass in order to have the appropriate time for inflation. In the following table we present our results for the observable quantities for a specific set of the parameter space $\left\{\alpha,\lambda,\beta\right\}$.

\begin{table}[H]
    \centering
    \begin{tabular}{||c||c||}
    \hline
         $\beta$& $n_s\ (N=55)$ \\\hline
         $0.019$   & $0.9761$     \\
         $0.020$   & $0.9721$       \\
         $0.021$   & $0.9681$       \\
         $0.022$   & $0.9641$        \\
         $0.023$   & $0.9601$       \\
         $0.024$   & $0.9561$       \\\hline
    \end{tabular}
    \caption{Values of the spectral tilt $n_s$ for $\alpha=10^{7}$, $\lambda=10^{-13}$ and varying values of $\beta$. Note that as $\beta$ increases the spectral tilt $n_s$ decreases rapidly, while the tensor-to-scalar ratio $r$ is largely unaffected, being $r_{N=55}\sim5\times10^{-3}$.}
    \label{table1}
\end{table}
The choice of the parameters\footnote{If the quartic model is to be identified as the Higgs, the issue of the metastability of the Higgs potential and the descent of the Higgs coupling to zero or even negative values should be addressed too. This issue is beyond the scope of the present article and the very small values of $\lambda$ are assumed on phenomenological grounds.} $\alpha=10^7$ and $\lambda=10^{-13}$ are such that they reproduce the observable value of the scalar power spectrum, as was also noted in \cite{Tenkanen2019, Gialamas2019, Tenkanen2020a}. If we allow for the case of varying $\alpha$, while keeping the other two parameters $\beta$ and $\lambda$ at constant values, we obtain the following table.
\begin{table}[H]
    \centering
    \begin{tabular}{||c||c||c||}
    \hline
         $\alpha$& $n_s\ (N=55)$ & $r\ (N=55)$\\\hline
         $5\times10^{6}$ & $0.9589$    	& $9.7\times10^{-3}$	\\
         $7\times10^6$   & $0.9639$      & $7.4\times10^{-3}$	 \\
         $9\times10^6$   & $0.9670$      & $5.9\times10^{-3}$	\\
         $10^7$   		& $0.9681$      & $5\times10^{-3}$ \\
         $2\times10^7$   & $0.9739$      & $3\times10^{-3}$		\\
         $3\times10^7$   & $0.9763$      & $2.1\times10^{-3}$				\\\hline
    \end{tabular}
    \caption{Values of the spectral tilt $n_s$ and the tensor-to-scalar ratio $r$ for $\beta=0.021$, $\lambda=10^{-13}$ and varying values of $\alpha\in\left\{5,30\right\}\times10^6$.}
    \label{table2}
\end{table}
For increasing values of $\alpha$ we expect an increase in the scalar spectral index $n_s$ and a decrease in the tensor-to-scalar ratio $r$. It is known, that in the slow-roll approximation, i.e. $\beta\to0$, the spectral index $n_s$ is explicitly independent of $\alpha$~\cite{Enckell2018a, Antoniadis2018, Antoniadis2019b}, at least without accounting for the subleading contribution of the additional kinetic term $B(\phi)$. In the present case of constant-roll, we note a dependence of $n_s$ on the values of $\alpha$ as well as the, already known, decrease in $r$ as $\alpha$ increases. It is worth mentioning that all the values of $\alpha$, presented in the above table, together with the values of $\beta$ and $\lambda$, produce a power spectrum $P_S\sim 10^{-9}$, close to its observed value. The results are well within the allowed region of $n_s$ and $r$ from the Planck2018 collaboration~\cite{Akrami2018} (see also \cite{Ade2018}), which is
\begin{equation}
    n_s=\left\{\begin{tabular}{c c}
          $\left(0.9607,0.9691\right)$,&$1\sigma\ \text{region}$ \\
          $\left(0.9565,0.9733\right)$,&$2\sigma\ \text{region}$
    \end{tabular}\right.,\qquad r<0.064.
\end{equation}
Note that the same $R^2$--Palatini quartic model, considered in the framework of slow-roll inflation, required an unusually large amount of $e$-folds ($N\sim75$) in order to agree with the observational bounds on $n_s$. This is in fact recovered taking the slow-roll limit $\beta\to0$. In contrast, in the present case of constant-roll inflation with $\beta\sim10^{-2}$, we obtain the desired results for the $n_s$ and $r$ with the appropriate amount of $e$-folds ($N\,\approx\,50-60$). It is also worth noting that during inflation we have the following order of magnitude hierarchy of the kinetic terms:
\begin{equation}
    \frac{AX}{BX^2}\sim\left\{\begin{matrix}10^{-2},&\phi=\phi_i\\1,&\phi=\phi_f\end{matrix}\right.,\ \text{at }N=50,\qquad\qquad\qquad \frac{AX}{BX^2}\sim1,\ \text{at }N=60,
\end{equation}
for $\alpha=10^7$, $\lambda=10^{-13}$ and $\beta=0.021$ . This shows clearly that the additional quartic kinetic term has a considerable contribution to the overall dynamics of the inflaton field, when the constant-roll condition is applied. It is important to note that the assumptions made for the smallness of the slow-roll parameters $\epsilon_i\ll1$, made for instance in order to calculate the tensor-to-scalar ratio, are valid throughout. For example, in the case  $\alpha=10^{7}$, $\lambda=10^{-13}$ and $\beta=0.021$, we obtain $\epsilon_1\simeq10^{-3},\,\,\,\left|\epsilon_4\right|\sim 10^{-2}$. Additionally, it has been also verified numerically that the analytic solution of $\dot{\phi}[\phi]$ does satisfy the field equation of motion \eqref{EQONO-1}, with small deviations at field values after the end of inflation, $\phi<\phi_f$, as expected. Nevertheless, it should be noted that inflationary dynamics takes place in the large field domain as it becomes clear from the scale of inflation. The scale of inflation is determined in terms of the canonical field $\Phi$, defined as
\begin{equation}
-\frac{1}{2}\left(\nabla\Phi\right)^2=\frac{1}{2}\,A(\phi)\left(\nabla\phi\right)^2+\frac{1}{4}\,B(\phi)\left(\nabla\phi\right)^4,
\end{equation}
which, in the case of the specific minimal quartic model studied here, we can simplify it as
\begin{equation}\label{Phidef}
\left(\frac{\mathrm{d}\Phi}{\mathrm{d}\phi}\right)^2=-A(\phi)\left(1+\frac{\alpha}{2}\,\dot{\phi}^2\right)=\frac{1}{1+\alpha\lambda\phi^4}\left(1+\frac{\alpha}{2}\,\dot{\phi}^2\right).
\end{equation}
Next, we can substitute the solution $\dot{\phi}[\phi]$, expanding it first around large $\phi$ as $\dot{\phi}\simeq\vartheta_0+\vartheta_1/\phi$, where $\vartheta_i$ are constants depending only on the values of the model parameters $\alpha$, $\lambda$ and $\beta$. This is consistent with the values assumed by the field $\phi$ at the end and start of inflation, residing in the  transPlanckian region. Then, the relation between the initial field $\phi$ and the canonically normalized field $\Phi$ is, approximated for large $\phi$,
\begin{equation}
\Phi\simeq C\mp \left(\sqrt{\frac{1+\frac{\alpha}{2}\,\vartheta_0^2}{\alpha\lambda}}\right)\frac{1}{\phi}\mp\left(\frac{\alpha\,\vartheta_0\,\vartheta_1}{4\sqrt{\alpha\lambda}\sqrt{1+\frac{\alpha}{2}\vartheta_0^2}}\right)\frac{1}{\phi^2},
\end{equation}
up to an arbitrary constant $C$. Independently of $C$, one can calculate the exact field excursion $\Delta\Phi$ directly from \eqref{Phidef}. If we assume that $\alpha=10^7$, $\lambda=10^{-13}$ and $\beta=0.021$, we obtain a value of $\Delta\Phi\simeq14M_P$. Given the inverted relation between $\Phi$ and $\phi$, with the end of inflation at some subPlanckian value $\Phi_i\sim\mathcal{O}(10^{-2})M_P$, we obtain $\Phi_f\sim 14M_P$. Note that the field equation of motion \eqref{EQONO-1}, now expressed in terms of $\Phi$, has to be satisfied, at least in the transPlanckian $\phi$ domain reflecting the inflationary period. In the case discussed here, this has been confirmed.

\clearpage
{\textbf{Non-minimally coupled quartic potential.}} Since, quantum corrections are expected, in addition to an $R^2$ term, to induce a $\xi\phi^2R$ term as well, we include in our analysis the case of $\xi\neq0$. In fact, it is known, that in the framework of slow-roll Palatini inflation, the interplay between this non-minimal coupling and an $R^2$ term allows for appropriate values for the inflationary observables for a small coupling constant $\xi$, in contrast to the metric formulation of Higgs inflation where large values of $\xi$ are necessary.

Following the same method of analysis, as in the previous case but for $\xi\neq0$, we present our predictions for the $n_s$ and $r$ observables in the next figure.

\begin{figure}[H]
    \centering
    \includegraphics[scale=0.5]{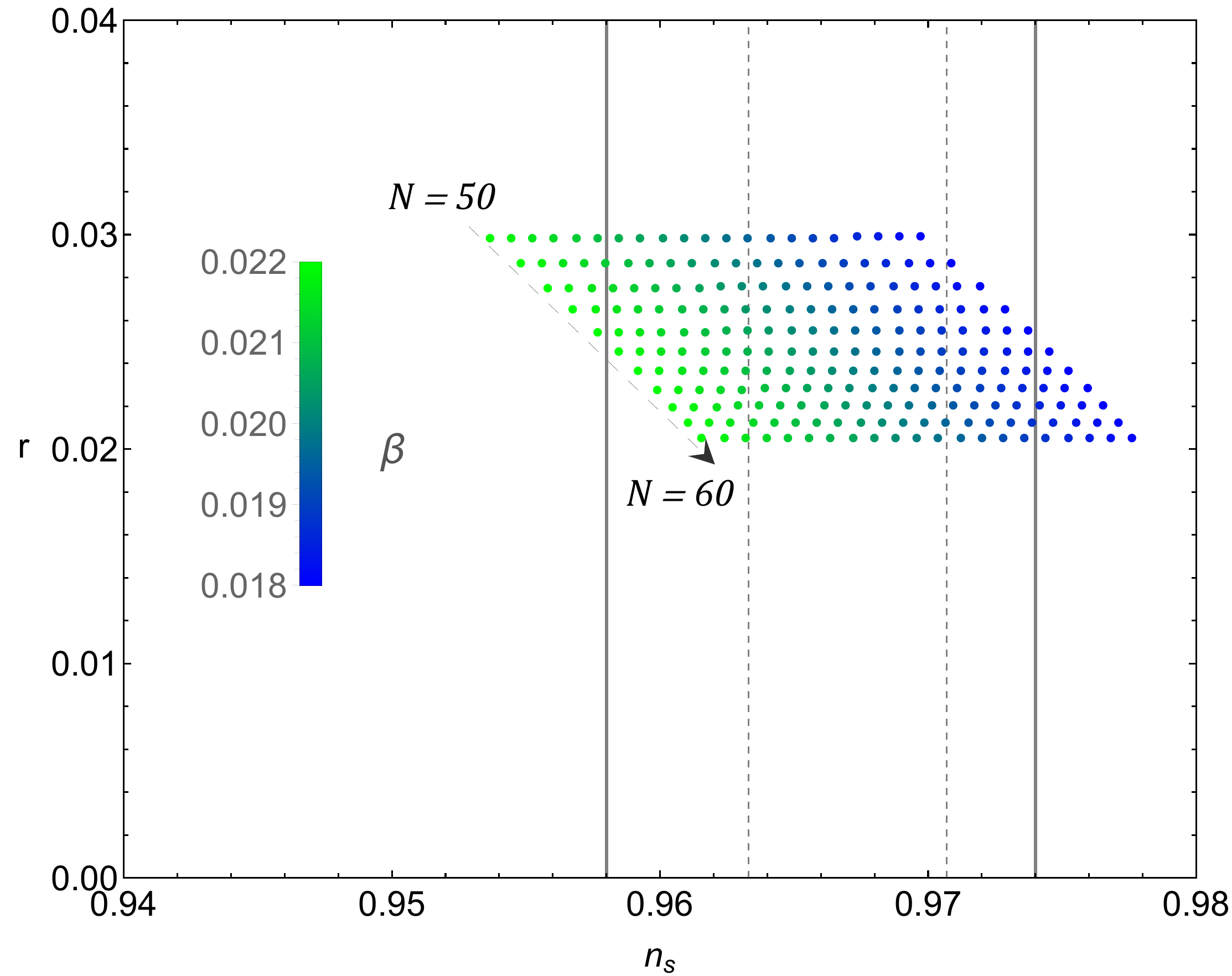}
    \caption{A plot of the $n_s-r$ plane, for $\alpha=10^7$, $\lambda=10^{-13}$, $\xi=10^{-6}$ and varying values of $\beta\in\left\{0.018,0.022\right\}$. The dashed and solid lines represent the $1\sigma$ and $2\sigma$ allowed range of the $n_s$, respectively, while the whole of the presented parameter space is allowed for $r$ ($r<0.06$). Once again, as $|\beta|$ increases, the spectral tilt $n_s$ decreases, while the effect on the tensor-to-scalar ratio $r$ is minimal, $r_{N=50}=3\times10^{-2}$ and $r_{N=60}=2\times10^{-2}$.}
\end{figure}

The values of the parameters, $\alpha=10^7$, $\lambda=10^{-13}$ and $\xi=10^{-6}$ are chosen such that they reproduce the observable value of the power spectrum in \eqref{powerspectrum}. As expected the model can provide an appropriate amount  of inflation in the large field domain with relevant results for the inflationary observables. Also, we find that the slow-roll parameters indeed assume small values, in this case as well, i.e. $\epsilon_1\lesssim10^{-2}$ and $|\epsilon_4|\sim10^{-2}$.

The numerical value of the parameter $C_s^2$, both in the case of minimal and non-minimal coupling, is $C_s^2\simeq0.34$. As expected we obtain values of $C_s^2<1$, while also avoiding instabilities associated with $C_s^2<0$. This value holds true throughout the stages of inflation, however, when $\phi\sim M_P$, the parameter tends rapidly to $C_s^2\sim1$.

\section{Summary}

In the present article we investigated the inflationary/phenomenological predictions of a scalar field coupled to quadratic gravity in the framework of the Palatini formulation. The resulting Einstein-frame generalized $k$-inflation effective theory was analyzed assuming that a constant-roll condition $\ddot{\phi}\!\sim\!\beta H\dot{\phi}$ holds. The equations of motion were derived, as well as the general expressions for the slow-roll parameters and observational indices. We focused on a quartic self-interaction potential, a case of particular appeal modelling Higgs inflation, considering both the cases of minimal and non-minimal coupling of the inflaton to gravity. The results show a significant contribution from the additional kinetic terms, in contrast to the slow-roll analysis of the same model. For an appropriate range of the model parameters, with inflationary dynamics taking place in the large field domain,  we obtained values for the inflationary observables, namely the primordial tilt $n_s$, the tensor-to-scalar ratio $r$ and the power spectrum, in the general range allowed by observations both in the case of the minimal coupling as well as the case of non-minimal coupling of the scalar to gravity. Contrary to the slow-roll analysis of the same model, where these results required an unusually large amount of $e$-folds, in the present case of constant-roll inflation the appropriate amount of $e$-fold was retained.

\section*{Acknowledgements}
A.L. and K.T. thank Alexandros Karam for discussions. This work was supported in part by the Labex ``Institut Lagrange de Paris’' and in part by a CNRS PICS grant. The research of A.L. is co-financed by Greece and the European Union (European Social Fund - ESF) through the Operational Programme ``Human Resources Development, Education and Lifelong Learning" in the context of the project ``Strengthening Human Resources Research Potential via Doctorate Research'' (MIS-5000432), implemented by the State Scholarships Foundation (IKY). K.T. would like to thank the Albert Einstein Institute of Theoretical Physics for hospitality and financial support.

\bibliography{References}{}
\bibliographystyle{jhep}

\end{document}